\documentstyle[floats,prl,aps,epsf]{revtex} 
\begin{document}
\twocolumn[\hsize\textwidth\columnwidth\hsize\csname@twocolumnfalse\endcsname


\title{Comparing the inspiral of irrotational and corotational binary
	neutron stars}

\author{Matthew D. Duez${}^1$, Thomas W. Baumgarte${}^{1,2}$, 
	Stuart L. Shapiro${}^{1,3}$, Masaru Shibata${}^4$, 
	and K\=oji Ury\=u${}^5$}
\address{
${}^1$ 	Department of Physics, University of Illinois 
	at Urbana-Champaign, Urbana, IL 61801 \\
${}^2$ 	Department of Physics and Astronomy, Bowdoin College,
	Brunswick, ME 04011 \\
${}^3$  Department of Astronomy \& NCSA, University of Illinois 
	at Urbana-Champaign, Urbana, IL 61801 \\
${}^4$  Graduate School of Arts and Sciences, University of Tokyo, 
	Komaba, Meguro, Tokyo 153-8902, Japan \\
${}^5$  Department of Physics, University of Wisconsin-Milwaukee, 
	P.O. Box 413, Milwaukee, WI  53201 \\
}
\maketitle

\begin{abstract}
We model the adiabatic inspiral of relativistic binary neutron stars
in a quasi-equilibrium (QE) approximation, and compute the gravitational
wavetrain from the late phase of the inspiral.  We compare
corotational and irrotational sequences and find a significant
difference in the inspiral rate, which is almost entirely caused by
differences in the binding energy.  We also compare our results with
those of a point-mass post-Newtonian calculation.  We
illustrate how the late inspiral wavetrain computed with our QE
numerical scheme can be matched to the
subsequent plunge and merger waveform calculated with a fully
relativistic hydrodynamics code.
\end{abstract}

\draft 

\vskip2pc]

\section{Introduction}

Binary neutron stars are among the most promising sources for the
detection of gravitational waves by the laser interferometers
currently under development, including LIGO, VIRGO, GEO and TAMA
(see~\cite{tetal01} for the first preliminary data analysis from the
TAMA detector).  Typical binary neutron stars are expected to complete
about 16,000 orbital periods (corresponding to about 15 minutes) while
sweeping through LIGO's frequency band between 10 and 1000
Hz~\cite{t98}.  Accurate knowledge of the binary inspiral and the
associated gravitational waveform is crucial to enhance our chances of
detecting the signal in the noisy output of the laser interferometer.

The evolution of binary neutron stars proceeds in different stages.
By far the longest is the initial quasi-equilibrium {\it inspiral} phase,
during which the stars move in nearly circular orbits, while the
separation between the stars decreases adiabatically as energy is carried
away by gravitational radiation.  The quasi-circular orbits become
unstable at the innermost stable circular orbit (ISCO), where the
inspiraling system enters a {\it plunge and merger} phase.  The merger and
coalescence of the stars happens on a dynamical timescale, and
produces either a black hole or a larger neutron star, which may
collapse to a black hole at a later time.  The final stage of the
evolution is the {\it ringdown} phase, during which the merged object
settles down to equilibrium.           

The early inspiral phase, for large binary separations, can be modeled
very accurately with post-Newtonian (PN) methods, e.g.~\cite{bdiww95}.
Recently, the convergence of these PN expansion has been improved by
the introduction of Pad\'e approximants~\cite{dis98}, and an
alternative PN ``effective one-body'' approach has also been
suggested~\cite{bd99} (see also~\cite{djs00} and references therein).
It is generally accepted that the plunge and merger phase has to be
simulated by means of a self-consistent, fully relativistic hydrodynamics
calculation (see~\cite{su00} for the first such simulation).  During
the late ringdown phase, the merged object can be approximated as a
distorted equilibrium object, so that perturbative techniques can be
applied (e.g.~\cite{pp94,ketal99}).

It is likely, however, that PN point-mass techniques break
down somewhat outside of the ISCO, when finite-size and relativistic
effects become important.  It is hard to imagine that fully hydrodynamical
numerical calculations will be able to follow the inspiral reliably 
by beginning far beyond that
point, through many orbital periods, all the way to the ISCO, followed
by plunge and
merger.  Such calculations would accumulate significant amounts
of numerical error and would be computationally prohibitive.  This
leaves a gap between the regimes that
PN and fully numerical calculations can model.  Filling
this gap for the late inspiral, immediately prior to plunge, therefore
requires an alternative approach (in the case of binary black holes,
this problem has been called the ``intermediate binary black hole''
problem~\cite{bct98}).

We have recently adopted a quasi-equilibrium (QE) approach to model
the late inspiral~\cite{dbs01}.  This approach takes advantage of the
fact that during the inspiral phase, the binaries evolve slowly on nearly
circular orbits.  The matter distribution can therefore be computed
independently by assuming the stars to be in quasi-equilibrium
(e.g.~\cite{bcsst97,ue00}).  These pre-determined matter profiles,
which satisfy the hydrodynamic equations of quasi-static equilibrium,
can then be inserted into Einstein's field equations as source terms
and the hydrodynamic equations do not have to be
solved again (see the ``hydro-without-hydro'' approach discussed
in~\cite{bhs99}).  For a given separation, we evolve the binary for
about two orbital periods to determine the gravitational wave form
and luminosity.  We repeat the calculation for a discrete set of
separations, which then allows us to construct the entire late
inspiral together with its gravitational wave signal.  The QE method
is described in more detail in~\cite{dbs01} and in Section~\ref{sec2.1}
below.  We have
also tested and calibrated this method for a model problem in relativistic
scalar gravitation~\cite{ybs01}, where the QE wavetrain could be compared
with the full wavetrain calculated by an exact numerical integration. 

In~\cite{dbs01} we presented a prototype calculation for a
corotational binary sequence.  It is unlikely, however, that binary
neutron stars can maintain corotation during inspiral~\cite{irr_1}.
In this paper, we therefore compare with the more realistic case of
irrotational binary neutron stars.  By comparing the results we thereby
explore the sensitivity of the wavetrain to the internal fluid
velocities.  We discuss the criterion which determines the point of
breakdown of the QE approximation just outside the ISCO.  We thus
establish the innermost separation at which a QE calculation can
provide initial
data for future dynamical simulations of the plunge and merger.
Finally, we compare our computational results for the late inspiral
with a point-mass PN calculation of the same epoch.

The paper is organized as follows.  In Section~\ref{sec2} we describe
our numerical methods and resulting waveforms (Section~\ref{sec2.1}), 
present a first order PN
formalism (Section~\ref{sec2.2}), explain the construction of an
inspiral wavetrain (Section~\ref{sec2.3}), and finally compare
corotational and irrotational sequences, in both the numerical and PN
approach (Section~\ref{sec2.4}).  In Section~\ref{sec3} we illustrate
how an entire wavetrain, spanning both the late inspiral and the plunge and
merger phases, can be constructed by matching QE and fully dynamical
results.  We briefly summarize our results in Section~\ref{sec4}.

\section{Irrotational and corotational binary inspiral}
\label{sec2}

\subsection{Numerical results}
\label{sec2.1}

\begin{table*}[t]
\caption{\normalsize{Irrotational Sequence}}
\begin{tabular}{ccccccccc}
 $\quad z_A{}^a\quad$ & $\quad M/M_0{}^b\quad$ & $\quad\Omega_{\rm orb} M_0\quad$
&  $\quad A M_0{}^c\quad$  &  $\qquad dM/dt\qquad$  & $\quad r/M_0{}^d \quad$
& $\qquad dr/dt\qquad$  & $\quad N_{\rm cyc}{}^e\quad$  &
$\qquad\xi\qquad$ \\ \hline                            0.500 & 0.95261
& 0.00587 & 0.0190 & 4.94e-09 & 14.7 & 2.21e-05  &  0.0
& 0.001582 \\ 0.474 & 0.95237 & 0.00648 & 0.0219 &
8.022e-09 & 13.7 & 2.79e-05  & 77.1 &
0.00250 \\ 0.444 & 0.95210 & 0.00722 & 0.0234 & 1.14e-08 &
12.7 & 3.73e-05  & 144.9 & 0.00351 \\ 0.412 &
0.95177 & 0.00808 & 0.0261 & 1.78e-08 & 11.8 & 5.12e-05  &
198.3 & 0.00552 \\ 0.375 & 0.95141 & 0.00913 &
0.0290 & 2.80e-08 & 10.8 & 7.37e-05  & 242.4 &
0.00898 \\ 0.333 & 0.95100 & 0.0104 & 0.0335 & 4.84e-08 &
9.89 & 0.000110  & 277.0 & 0.0168 \\ 0.286 &
0.95052 & 0.0119 & 0.0379 & 8.20e-08 &  8.99 & 0.000169  &
302.8 & 0.0343 \\ 0.231 & 0.95001 & 0.0138 &
0.0440 & 1.47e-07 &  8.10 & 0.000277  & 322.1 &
0.0896 \\ 0.167 & 0.94949 & 0.0159 & 0.0525 & 2.79e-07 &
7.28 & 0.000563  & 334.9 &  0.483 \\ $\ast$ 0.130 &
0.94929 & 0.0169 & 0.0569 & 3.69e-07 &  6.91 &  0.00135 &
338.5 & 4.90 \\ $\ast$ 0.111 & 0.94926 & 0.0170 &
0.0580 & 3.87e-07 &  6.78 &   0.0121  & 339.1 &
$\rm\infty$ \\
\end{tabular}

${}^a$ $z_A$ is defined to be the ratio of the separation between the 
innermost points and the outermost points on the stars.   

${}^b$ $M_0$ is the total rest mass of the system, i.e., 
twice the rest mass of an individual star.

${}^c$ $A$ is the amplitude of gravitational waves on the rotation axis. 

${}^d$ $r$ is defined as the average of the coordinate distance from the 
origin of the nearest and farthest points on a neutron star.  
Thus, it is the coordinate radius of the ``center'' in 
$\kappa=1$ units and the metric and coordinate system of 
\cite{bcsst97,ue00}.

${}^e$ $N_{\rm cyc}$ is the number of gravitational wave cycles from 
our initial configuration (i.e. that of largest $r$) to the 
indicated separation.

$\ast$ QE approximation is invalid:  $\xi > 1$.

\end{table*}

\begin{table*}[t]
\caption{\normalsize{Corotational Sequence}}
\begin{tabular}{ccccccccc}
 $\quad z_A\quad$ & $\quad M/M_0\quad$ & $\quad\Omega_{\rm orb} M_0\quad$
&  $\quad A M_0\quad$  &  $\qquad dM/dt\qquad$ & $\quad r/M_0 \quad$
& $\qquad dr/dt\qquad$  & $\quad N_{\rm cyc}\quad$ &
$\qquad\xi\qquad$ \\ \hline                            0.475 &
0.952098 & 0.00656 & 0.0217 & 8.06e-09 & 13.7  & 4.57e-05
&  0.0  & 0.00343 \\ 0.45 & 0.95190  & 0.00715 &
0.0236 & 1.13e-08 & 12.9  & 3.66e-05  & 46.6
& 0.00486 \\ 0.400 & 0.951493 & 0.00843  & 0.0270 &
2.07e-08 & 11.4 & 6.40e-05  & 128.1 &
0.00938 \\ 0.375 & 0.95131  & 0.00922  & 0.0299  & 3.04e-08 &
10.8  & 9.32e-05  & 152.0 & 0.01416 \\ 0.350 &
0.951106 & 0.0102 & 0.0341  & 4.82e-08 & 10.1  & 0.000130
& 168.8 & 0.0236 \\ 0.300 & 0.950752 & 0.0114 &
0.0368  & 7.06e-08  &  9.27 & 0.000242 & 189.8
& 0.0429 \\ 0.275 & 0.950569 & 0.0122 & 0.0393 & 9.20e-08 &
8.84 & 0.000320  & 196.5 & 0.0656 \\ 0.250 &
0.950471 & 0.0130 & 0.0415  & 1.16e-07  &  8.45 & 0.000422
& 201.6 & 0.09014 \\ 0.225 & 0.950329 & 0.0138 &
0.0446  & 1.52e-07 &  8.09 & 0.000557  & 205.5
& 0.143 \\ 0.200 & 0.950228 & 0.0146 & 0.0477  & 1.93e-07 &
7.78  & 0.000744  & 208.4 & 0.219 \\ 0.175 &
0.950093 & 0.0154  & 0.0500 & 2.36e-07 &  7.47 & 0.00101
& 210.6 & 0.393 \\ 0.15 & 0.950031 & 0.0161 &
0.0540   & 3.02e-07 & 7.21 & 0.00140 & 212.2
& 0.645 \\ 0.125 & 0.950024 & 0.0168 & 0.0565  & 3.59e-07 &
6.98 & 0.00204 & 213.3 & 0.761 \\ $\ast$ 0.100 &
0.949935 & 0.0174 & 0.0595 & 4.31e-07 &  6.75  & 0.00312
& 214.1 & 1.77 \\
\end{tabular}
\end{table*}

We adopt a quasi-equilibrium (QE) approximation to study the adiabatic,
late inspiral of binary neutron stars, taking advantage of the fact
that in this regime the orbital decay timescale is
much longer than the orbital period, and, in addition, the gravitational
radiation has negligible effect on the structure of the stars 
(see~\cite{dbs01} for details).
In this approximation, the inspiral is modeled as a sequence of
``snapshots'' of binaries in quasi-circular orbit.  In this paper we
generalize the results of~\cite{dbs01} for corotational binaries
(based on the models of~\cite{bcsst97}) to irrotational binaries
(based on the models of~\cite{ue00}) and analyze the effect of
spin on the inspiral.  The snapshots used were constructed
under the assumptions that the metric is asymptotically and
conformally flat, although the metric ceases to be conformally flat
as we evolve it to generate waveforms.  These models provide the
data to construct binding energy curves, which give the total
mass-energy $M(r)$ as a function of separation $r$ for circular binaries
of fixed rest (baryonic) mass.

We use these binary models to provide the matter source terms
in the Einstein field equations, which we integrate numerically with
the formalism and code described in~\cite{bs99} (see also~\cite{sn95})
and used in~\cite{dbs01}. 
For binary separations where the QE approximation is valid,
the hydrodynamic equations do not need to be resolved after the initial
time .  Instead, we freeze the matter variables at their equilibrium
values and move the stars on circular orbits while allowing the metric to
evolve dynamically.  We
typically evolve the binary for about two orbital periods for each
orbital separation, which is sufficient to read off the gravitational waveform
$h$ and the gravitational wave luminosity $-dM/dt$.  The initial
gravitational wave
signal is corrupted by noise contained in the approximate initial
data, which quickly propagates off the numerical grid leaving
a periodic wave signal from the binary.  Repeating the
integration for various binary separations $r$ yields $h(r)$ and
$dM/dt(r)$ as a function of separation.  The inspiral rate can then be
found by combining the luminosity with the derivative of the binding
energy with respect to separation,
\begin{equation} \label{inspiral_rate}
\frac{dr}{dt} = \frac{dM/dt}{dM/dr}.
\end{equation}
Integrating this equation yields the separation $r$ as a function of
time $t$.  The gravitational wavetrain $h(t)$ can finally be
constructed by using $r(t)$ to fit the discrete waveforms $h(r)$
together in a smooth and continuous fashion.  This QE approach has
also been tested and calibrated for a model problem in relativistic
scalar gravitation (\cite{ybs01}).

 Both the corotational and the irrotational sequences terminate at some
innermost circular orbit (ICO), at which the quasi-equilibrium
approximation breaks down.  For the corotational sequences, this ICO
corresponds to the innermost stable circular orbit (ISCO) at which the
equilibrium orbits
become unstable, and at which the plunge and merger of the binary set
in.  Typical irrotational equilibrium sequences terminate before they
reach the
ISCO, when they form cusp-like structures at the inner edges of the
stars~\cite{ue00,use00}.  The cusps indicate the onset of mass overflow. 
The termination of the equilibrium sequence indicates the break-down
of the QE approximation, since no equilibrium object exists with this
separation.  For both the corotational and the irrotational sequences,
we denote the mass-energy at the innermost circular orbit as $M_{\rm ICO}$.

By adopting QE matter sources in our evolution
calculations, we assume that these remain stationary in a co-rotating
frame of reference over an orbital period.  As a necessary
condition for this to be true, we have to ensure that the binary 
does not reach the ICO in approximately one orbital period $P$.  Since
the energy emitted by the binary per period is $(dM/dt)P$, this condition
holds as long as the dimensionless ratio
\begin{equation}
\label{xidef}
\xi \equiv \frac{|dM/dt| P}{M - M_{\rm ICO}}
\end{equation}
is less than unity.  Identifying the orbital timescale $t_{\rm orb}$
with $P$ and the gravitational wave inspiral timescale $t_{\rm GW}$ with
$M/|dM/dt|$ shows that the condition $\xi < 1$ is identical to
\begin{equation}
\label{ineq}
\frac{M - M_{\rm ICO}}{M} > \frac{t_{\rm orb}}{t_{\rm GW}}.
\end{equation}
Eq.~(\ref{ineq}) states that, for infinitesimally small 
$t_{\rm orb}/t_{\rm GW}$, we can apply the QE
approximation arbitrarily close to the ICO.  However, if $\xi > 1$,
the binaries will overshoot the innermost equilibrium orbit from their
current separation in less than one orbital period, causing a
breakdown in quasi-equilibrium.

The above criterion is not only useful for determining the limits of
the QE approximation, but also for choosing the separation for assigning
initial data for dynamical simulations of the plunge and merger. 
In order to reduce
computational resources and accumulation of numerical error, one would
like to impose such initial data as close to the ICO as possible.
Very close to the ICO, however, the approximations of the QE approach
break down, as indicated by Eq.~(\ref{xidef}).  The separation at which $\xi
\approx 1$ is therefore a suitable compromise for terminating the QE
approach, and imposing initial data for dynamical simulations.

In Tables I and II we summarize our numerical results for irrotational
and corotational binaries.  For both cases the stars are modeled as
polytropes with pressure $P = \kappa\rho_0^{1+1/n}$, where $\rho_0$ is
the rest-mass density, $n$ is the polytropic index, and $\kappa$ is a
constant.  We set $n=1$ and
non-dimensionalize our results by setting $\kappa = G = c = 1$.  In
these units, the individual stars have a rest mass $m_0 = 0.1$, which
is about 55\% of the maximum allowed rest mass of isolated,
spherical nonrotating stars, and corresponds to a compaction of
$m_{\rm \infty}/R_{\rm \infty} = 0.088$ in isolation.  Here $m_{\rm \infty}$ and
$R_{\rm \infty}$ are the mass-energy and areal radius (i.e. radius defined
by the area of a constant-$r$ shell) of the individual
(spherical) stars at infinite binary separation.  Our results apply to
arbitrary mass stars with this compaction and equation of state.  For
both cases, we study models up to and past $\xi = 1$.
In Figure 1, we verify that our evolution code satisfies the Hamiltonian
and momentum constraints.  This is a nontrivial test of our method:  The
matter profiles are determined by solving the constraint equations in the
QE approximation.  The fact that they remain satisfied during the field
evolution demonstrates that the spacetime remains close to
quasi-equilibrium, even as gravitational radiation is generated and the
metric is determined dynamically.

\begin{figure}
\epsfxsize=3.0in
\begin{center}
\leavevmode \epsffile{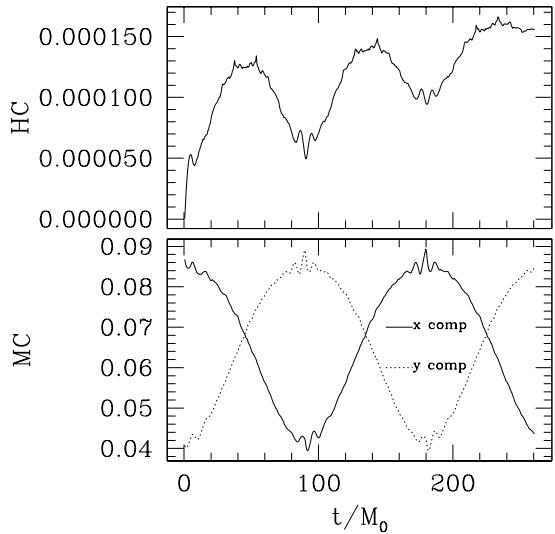}
\end{center}
\caption{ The L2 norm of the Hamiltonian constraint and of the
x and y components of the momentum constraint for corotating binaries
with separation parameter $z_A = 0.1$ orbiting about the $z$-axis. 
The Hamiltonian constraint is normalized to the
initial value of $16\pi M$, the momentum constraint to the L2 norm of 
the initial value of $8\pi S_x$, where $S_x$ is the dominant component
of the initial matter current. 
The L2 norm is defined as $L2(f) = \sqrt{\sum_i f_i^2}$ }
\end{figure}

\subsection{Post-Newtonian treatment}
\label{sec2.2}

We adopt the formalism developed in~\cite{w94} to study the binary
inspiral in a first order PN approximation (note that PN techniques
have been extended to at least third order; see, e.g.~\cite{djs00}).
In particular, we consider a system of two point masses in circular
orbit with binary separation $r$, total orbital angular momentum ${\bf
L_N}$, and total spin angular momentum ${\bf S}$.  In isolation, each
point mass has a mass-energy $m_{\rm
\infty}$.  From~\cite{w94} we then find the orbital frequency
\begin{equation}
\label{wpn}
\Omega_{\rm orb}^2 = {\mu\over r^3}\bigl(1 - {11\over 4}{\mu\over r} - {5\over
\mu r^2}{\bf L_N}\cdot{\bf S}\bigr),
\end{equation}
the total mass-energy
\begin{equation}
\label{Epn}
M = \mu + {\mu\over 4}\bigl(-{1\over 2}{\mu\over r}  + {27\over
	32}\bigl({\mu\over r}\bigr)^2  - {3\over 2r^2}{\bf L_N}\cdot{\bf
	S}\bigr),
\end{equation}
and the luminosity
\begin{eqnarray}
\label{dEpn}
\displaystyle {dM\over dt} & = & 
\displaystyle -{2\over 5}\bigl({\mu\over r}\bigr)^5  \bigl( 1 -
{379\over 42}{\mu\over r} + 4\pi\bigl({\mu\over r}\bigr)^{3/2} 
\nonumber \\
& & - {37\over
3}\bigl({\mu\over r}\bigr)^{3/2}{\bf L_N}\cdot{\bf S}  \bigr).
\end{eqnarray}
Here we identify
\begin{equation}
\label{mpn}
\mu = \biggl\{ \begin{array}{ll} 2 m_{\rm \infty} & \mbox{irrotational} \\ 
			2 (m_{\rm \infty} + {1\over 2}I\Omega_{\rm orb}^2) 
			& \mbox{corotational} 
		\end{array}
\end{equation}
for the total mass-energy of each star in isolation. 
To compare with the binaries of Section~\ref{sec2.1}, we choose
\begin{equation}
{\bf L_N} = r^2 \Omega_{\rm orb} {\bf e_z}
\end{equation}
for the total orbital angular momentum and 
\begin{equation}
\label{Spn}
{\bf S} = \biggl\{ \matrix{ {\bf 0} \ \qquad\mbox{irrotational} \cr
			    2 I\Omega_{\rm orb} {\bf e_z} \quad\mbox{corotational}
			    \cr }
\end{equation}
for the spin angular momentum.

In the above, $r$ is the separation of the point masses in harmonic
coordinates, and the moment of inertia of each mass, $I$, is calculated
for a slowly
rotating (nearly TOV) relativistic star \cite{ab99}.  Note that, for
the corotational case, $\mu$, and $\Omega_{\rm orb}$ must be solved
simultaneously to PN order.

\subsection{Constructing Wavetrains}
\label{sec2.3}

We extract gravitational wave data from our QE simulations by matching
the numerical data near the boundaries of our grids to the $l = 2$, $m =
\pm 2$ Moncrief~\cite{m74} variables $R_{22+}$ and $R_{22-}$ (see Appendix A
for details).  From these, the energy luminosity is given by
\begin{equation}
\label{dEvsR}
{dM\over dt} = {r^2\over 32\pi}\bigl[(\partial_{\rm t}R_{\rm 22+})^2
			+ (\partial_{\rm t}R_{\rm 22-})^2\bigr].
\end{equation}

To generate the wavetrain, we also need to determine the wave amplitude. 
Both the QE and PN models provide more information than just the
quadrupole amplitudes, but we have found \cite{dbs01} that the
amplitudes of other modes are much smaller than the $l = 2$,
$m = \pm 2$ modes, and we ignore them below.  The asymptotic amplitudes of
$h_+$ and
$h_{\times}$ for these modes are identical along the axis of rotation, and
are called $A$ below.  This amplitude is related to
the luminosity by the quadrupole relation
\begin{equation}
\label{Apn}
A = {1\over\Omega_{\rm GW}}\sqrt{10 {dM\over dt}},
\end{equation}
where $\Omega_{\rm GW} = 2\Omega_{\rm orb}$ is the frequency of the waves. 
This gives $A(r)$, which can be converted to $A(t)$ by solving
Eq.~(\ref{inspiral_rate}).  Then the complete wavetrain as seen by an
observer on the rotation axis at a distance $r_s$ from the source is given
by
\begin{equation}
\label{h_wt}
r_s h(t) = A(t)\cos(\int_0^t\Omega_{\rm GW}(t') dt')
\end{equation}

\subsection{Comparing irrotational and corotational inspiral}
\label{sec2.4}

\begin{figure}
\epsfxsize=3.0in
\begin{center}
\leavevmode \epsffile{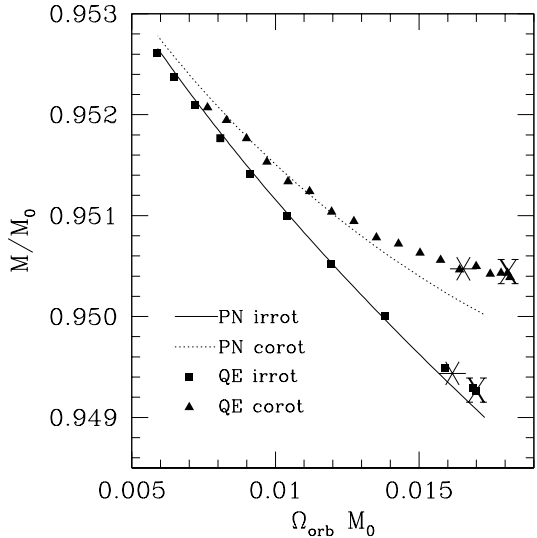}
\end{center}
\caption{The total mass-energy $M$ as a function of
orbital frequency for corotational and irrotational models. The cross
marks the ISO, the asterisk the location at which $\xi = 1$.}
\end{figure}

\begin{figure}
\epsfxsize=3.0in
\begin{center}
\leavevmode \epsffile{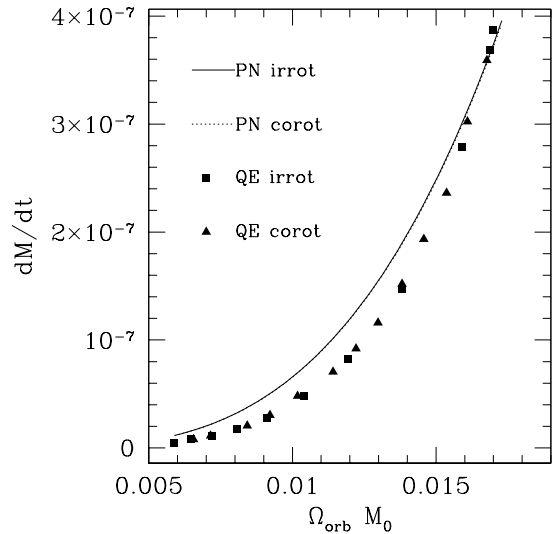}
\end{center}
\caption{The gravitational wave luminosity $dM/dt$ as a function of
orbital frequency for corotational and irrotational models. The solid
and dotted curves nearly coincide.}
\end{figure}

\begin{figure}
\epsfxsize=3.0in
\begin{center}
\leavevmode \epsffile{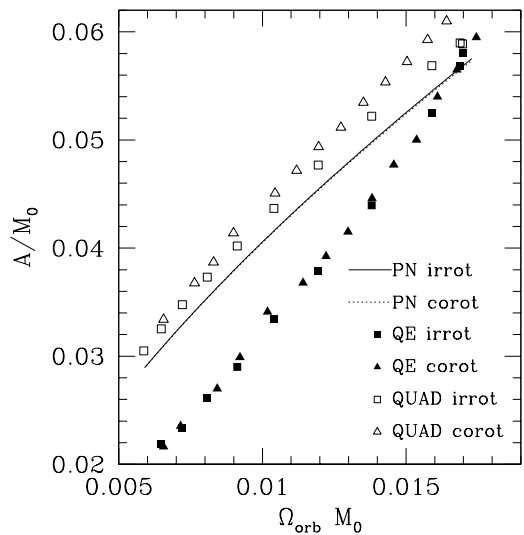}
\end{center}
\caption{Gravitational wave amplitude along the axis of rotation in
the QE, PN and Newtonian quadrupole approximation.
The PN curves for the two cases, which are based on Eq.~(\ref{Apn}), 
are hardly distinguishable.}
\end{figure}

\begin{figure}[t]
\epsfxsize=3.1in
\begin{center}
\leavevmode \epsffile{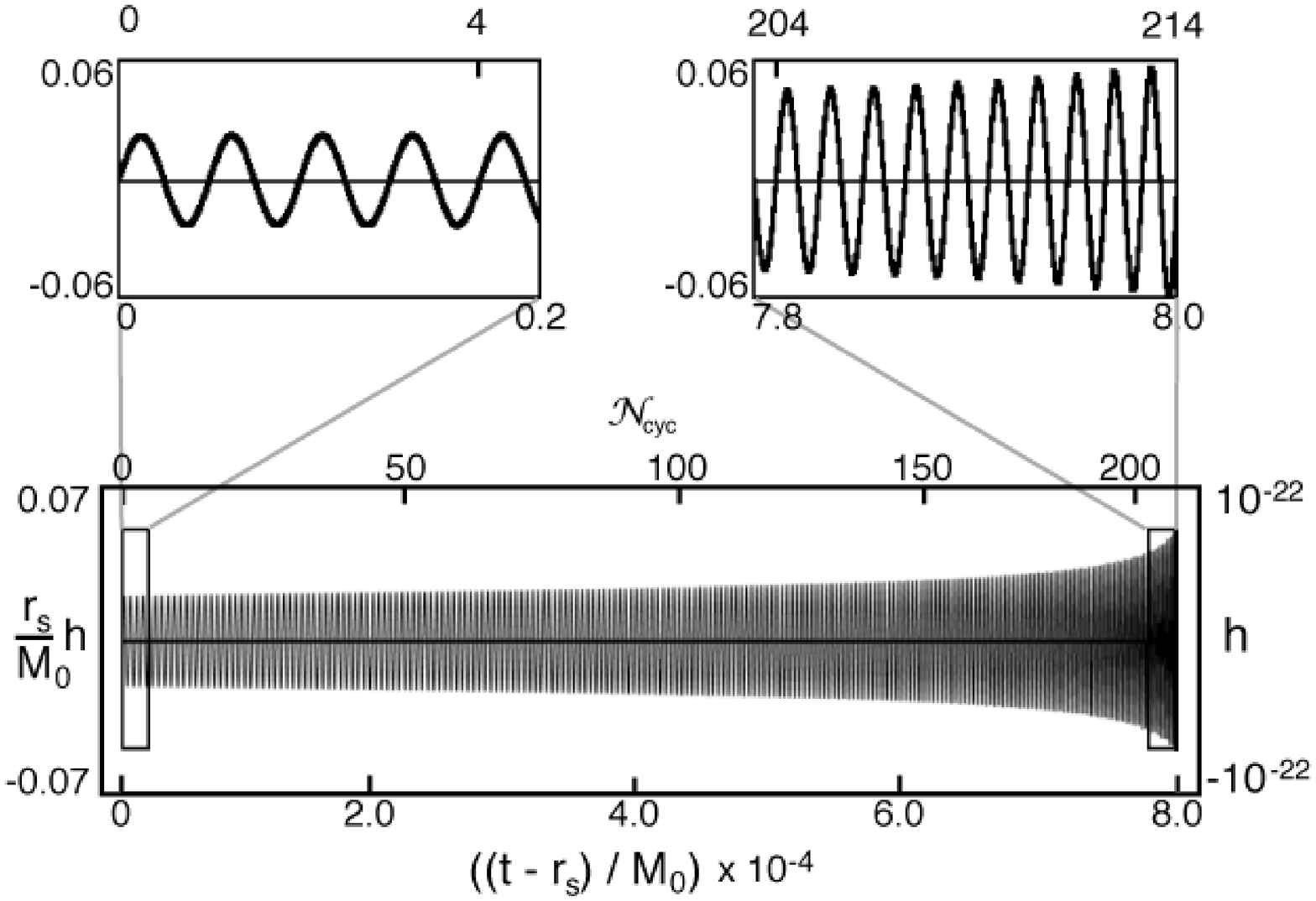}
\end{center}
\epsfxsize=3.1in
\begin{center}
\leavevmode \epsffile{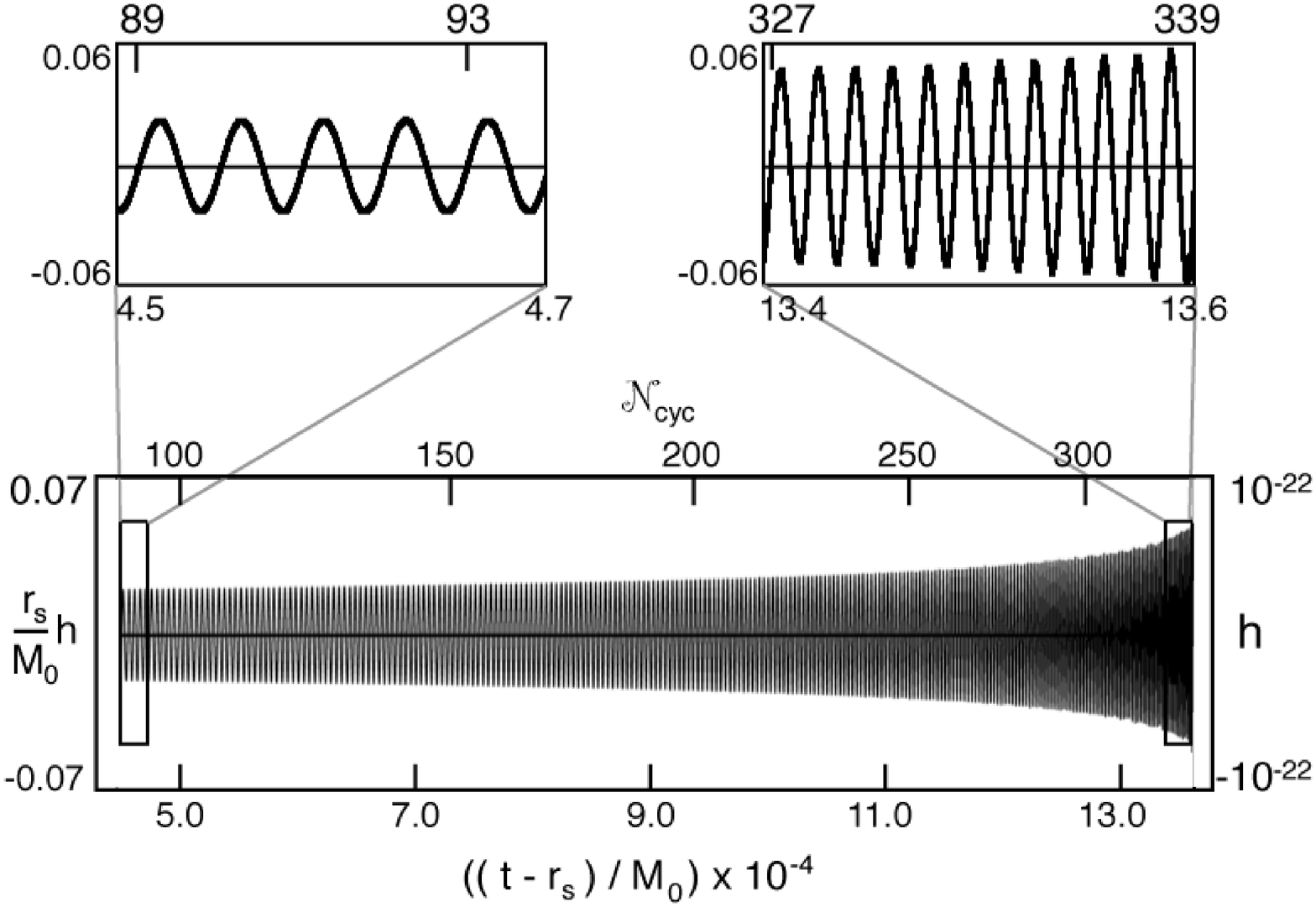}
\end{center}
\caption{
The final two hundred cycles of the inspiral waveform for
corotating (top) and irrotating (bottom) binaries. }
\end{figure}

In Figure 2, we show the total mass-energy $M$ of irrotational and
corotational inspiral sequences (along which the total rest mass $M_0$
is conserved).  We parameterize the binary separation by 
$\Omega_{\rm orb}$, the
orbital angular frequency as measured by a distant observer, since it
is an invariant. Note that infinite separation corresponds to 
$\Omega_{\rm orb}
= 0$.  The PN and QE results agree quite well for large separations,
while for small separations we find growing deviations between the two
approaches.  This is not surprising, since at small separations both
tidal interactions and relativistic effects play an increasingly
important role.  Finite size
effects are not taken into account by the point-mass calculation of
the PN approach.  The inclusion of 2 and 3PN terms will not
improve the agreement substantially, due to the dominance of finite
size effects~\cite{lrs94,b01}.  For small separations, the mass-energy,
and hence the binding energy, of irrotational and corotational sequences
is quite different.  This effect is due to the extra spin rotational
energy of the corotating binaries, which causes their total energy to
be larger than that of the irrotational binaries.  As we will see
below, this change in $M$, and hence $dM/dr$, dominates the
difference in the inspiral rate $dr/dt$.

In Figure 3, we plot $dM/dt$ for each model.  Note that the
gravitational wave luminosity $dM/dt$ is very similar for the
irrotational and corotational sequence in either the QE or the
PN approximation.  In Figure 4, we show the gravitational wave amplitudes
as measured along the axis of rotation and compare two different numerical
results with the PN values.  From our numerical binary models, we predict
the gravitational wave amplitude, both from the Newtonian quadrupole
approximation, where the quadrupole moments are computed by numerical
quadrature of the density and velocity fields, and by dynamically
evolving the (relativistic) gravitational fields and reading off the
wave amplitudes on the outer edge of the numerical grid, as described
in section IIC.

For large separations, the quadrupole and PN results agree quite well,
while the QE results differ by $\sim$ 30\%.  We attribute this
difference to the
inaccurate boundary treatment in our dynamical evolution
(see~\cite{dbs01}).  Gravitational radiation can only be read off
accurately in the far (wave) zone, at separations greater than a
gravitational wavelength $\lambda_{\rm GW}$.  In our simulations,
however, we are restricted to numerical grids which only extend to a
fraction of a gravitational wavelength, $r_{\rm max} \sim 0.1 - 0.35
\lambda_{\rm GW}$.  For large binary separations, the frequency
$\Omega_{\rm GW}$ is small, and hence $\lambda_{\rm GW}$ is large.  For
fixed outer boundary $r_{\rm max}$ this implies that
$r_{\rm max}/\lambda_{\rm GW}$
is smaller for larger binary separations, causing a more inaccurate
wave extraction.  As the separation becomes smaller, the QE and
quadrupole results agree increasingly well.
Simultaneously we expect the PN point-mass treatment to become
increasingly inaccurate for small separations, since it neglects
finite size effects.  This trend is reflected in the growing
deviations between the PN and quadrupole results in Figure 4.

The differences between the irrotational and corotational wave
amplitudes are very small in all three approaches.  This is quite
intuitive, since the gravitational wave emission is dominated by the
matter density distribution, which is fairly similar for the two
sequences, while matter current distributions play a less
important role.  The poor handling of the outer boundaries in the QE
approximation affects both sequences in the same systematic way, which
allows us to make a meaningful comparison.

The findings of this comparison with regard to the inspiral rate can be
anticipated from Eq.~(\ref{inspiral_rate}).  The numerator on the
right hand side, $dM/dt$ is very similar for irrotational and
corotational sequences, while the denominator, $dM/dr$ is different.
The latter therefore causes the inspiral rate $|dr/dt|$ to be smaller
for irrotational than for corotational binaries.  This effect
can be understood very easily even by a crude Newtonian argument, where
the total mass-energy of the star can be written as a sum
\begin{equation}
\label{NewtonE}
M \sim M_0 - \frac{M_0^2}{2r} 
	\ \left[ + 2 \frac{1}{2} I \Omega^2 \right] 
	\sim M_0 - \frac{M_0^2}{2r} 
	\ \left[ + I \frac{M_0}{r^3} \right].
\end{equation}
In Eq.~(\ref{NewtonE}), the first term on the right-hand side is the
energy of a spherical polytrope (rest $+$ internal $+$ gravitational
potential energy), the second term is the combined orbital kinetic and
gravitational potential energy for circular equilibrium, and the
third term, in square brackets, denotes the spin kinetic energy. 
Note that the term in the square brackets applies only to the corotational
sequence.  For a given separation, the energy of a corotational binary
is thus larger than that of an irrotational binary by the spin energy
of the individual stars.  Taking a derivative with respect to $r$ yields
\begin{equation}
\frac{dM^{\rm corot}}{dr} \sim \frac{dM^{\rm irrot}}{dr} - 3 I \frac{M_0}{r^4}.
\end{equation}
Inserting this into Eq.~(\ref{inspiral_rate}), and using the result
that $dM/dt$ is nearly identical for both sequences, immediately shows that
\begin{equation}
\left| \frac{dr}{dt} \right|_{\rm corot} >
\left| \frac{dr}{dt} \right|_{\rm irrot}.
\end{equation}
Correspondingly, the frequency of the gravitational wave train
increases at a faster rate for corotational binaries, which we
illustrate in Figure 5.

\section{Computing the full wavetrain through coalescence}
\label{sec3}

\begin{figure}[t]
\epsfxsize=3.0in
\begin{center}
\leavevmode
\epsffile{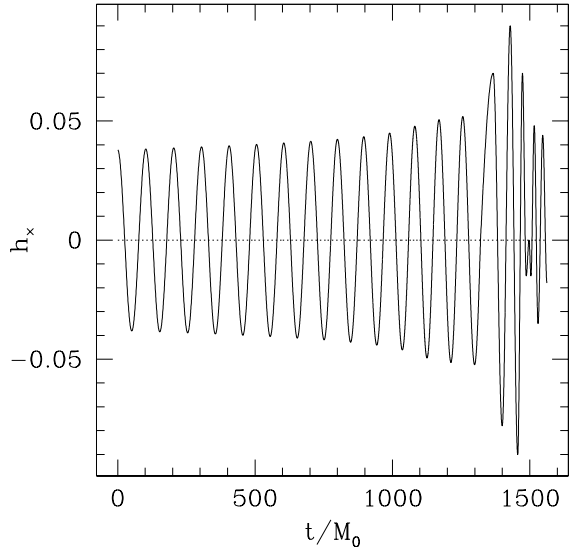}
\end{center}
\caption{A match of the total late QE inspiral wavetrain to the 
plunge and merger waveform, both constructed from numerical simulations
in general relativity.  Here we connect the last 13 orbits of the
inspiral, computed in the QE approximation, to the plunge calculation
of \protect\cite{su00}.  The ICO is reached at $t/M_0 = $1330, around
which point we pass from the adiabatic inspiral to the plunge and
merger phase. Here $h_{\times} = {r_s\over
M_0}\tilde\gamma_{\rm xy}$, where $r_s$ is the
distance from the source and $\tilde\gamma_{ij}$ is the conformal 
3-metric on the rotation axis.  This coalescence scenario leads to merger
and immediate black hole formation.  For $M_0 = 2 \times 1.5 M_{\odot}$
and $r_s = $100 Mpc,  the maximum amplitude of the metric perturbation
is $\sim 5 \times 10^{-21}$.}
\end{figure}

As we have argued in Section~\ref{sec2.1}, the QE approach could
provide accurate numerical initial data for fully relativistic hydrodynamics
simulations of the plunge and merger at a separation for which 
$\xi \begin{array}{rcl} &>& \\[-3mm] &\sim&\end{array} 1$. 
Starting at this separation guarantees that the
binary reaches the dynamical instability at the ICO
within about one orbital period.  This approach provide intrinsically
more accurate initial data than those imposed at the ICO (where the QE
approximation breaks down).  Moreover, the QE approach allows noise
contained in the initial data to be radiated away, thereby producing more
accurate equilibrium data.  In the QE approach, the initial noise
can be easily distinguished from the later periodic binary signal, whereas
in a dynamical plunge simulation it may be more difficult to distinguish
the propagation of initial noise from the intrinsically short-lived
coalescence waveform.

In Figure 6, we provide an illustration of what the combined late
inspiral-plunge wavetrain might look like.
We compute the late inspiral wavetrain using the QE method for an
irrotational binary sequence with stars of rest mass $m_0 = 0.146$,
which corresponds to 81\% of the maximum allowed rest mass of an
isolated, nonrotating star, and to a compaction of
$m_{\rm\infty}/R_{\rm\infty} = $0.14 in isolation~\cite{note}.  We then
match this to the full hydro dynamical waveform for the plunge and merger
of this same binary as computed by
Shibata and Ury\=u \cite{su00,note2}.  In order to get a continuous
waveform, we adopt the same wave extraction rules as in \cite{su00},
namely, $h_{\times}$ is read off the rotation axis at the edge of the
numerical grid, located at $r = 0.35
\lambda_{\rm GW}$.  The continuous wavetrain is constructed in the following
manner:  From $t = 0$ to $t = 1321 M_0$ at the ICO, we use QE
data.  A smooth (sine) function with the appropriate gravitational
wave frequency and amplitude is used to match the QE data to the
plunge data.  The raw plunge data given in~\cite{su00} begins at
$t = 1367 M_0$.

The location and nature of the transition from quasi-periodic inspiral
to rapid plunge is of great astrophysical interest because it is
sensitive to details of the binary neutron star system, in particular
the equation of state and rotation rate \cite{cabffkmopst93}.  As
illustrated above, the QE method, in conjunction with a relativistic
hydrodynamics code, could allow one to obtain the complete wavetrain
from late inspiral through plunge and merger, and at a very reasonable
computational cost.

\section{Summary}
\label{sec4}

We study the late inspiral of binary neutron stars and, by comparing
corotational and irrotational sequences, explore the effect of
internal fluid motions on the inspiral rate.  We find the
gravitational wave luminosity is very similar for any given radius.
However, the inspiral of corotational binaries proceeds faster.  

We also compare our numerical QE results with a first-order PN
calculation.  For large binary separations, the poor outer boundary
conditions in the numerical calculations introduce significant
deviations, while for small separations finite-size effects, which are
not taken into account in the PN calculations, can no longer be
neglected.  Future improvements in wave extraction and outer
boundary conditions (see, e.g.~\cite{outer_BC}) should reduce the
discrepancy.  These improvements remain a long-term goal for
numerical simulations of binary inspiral.

Lastly, we illustrate how QE and fully dynamical results could be
matched to produce the entire gravitational wavetrain,
covering the late inspiral through the plunge and merger phases. 
We emphasize that this paper presents only a prototype calculation showing
how, in principle, the continuous wavetrain can be assembled from
numerical simulations.

\acknowledgments
We would like to thank REU students J. Mehl, H. Agarwal,
P. McGrath, and D. Webber for assistance on Figure 5.  TWB gratefully
acknowledges financial support through a Fortner Fellowship at the
University of Illinois at Urbana-Champaign (UIUC).  Much of the
calculation and visualization were performed at the National Center
for Supercomputing Applications at UIUC.  This paper was supported in
part by NSF Grant PHY 99-02833 and NASA Grant NAG 5-10781 at UIUC and
by NSF Grant PHY00-71044 at the University of Wisconsin.

\begin{appendix}

\section{Gravitational wave extraction}

The two gravitational wave polarizations can be expressed as 
\begin{equation}
\begin{array}{rcl}
h_+ &=& \displaystyle 
	{{r}\over{2}}(\gamma^{TT}_{\rm \hat\theta \hat\theta} 
	- \gamma^{TT}_{\rm \hat\phi \hat\phi}) \\[3mm]
h_{\times} &=& r \gamma^{TT}_{\rm \hat\theta \hat\phi},
\end{array}
\end{equation}
where the superscript TT denotes transverse-traceless projection, and
$\gamma_{\rm \hat i \hat j}$ is the spatial metric in orthonormal
spherical-polar coordinates.  Note that the amplitudes of $h_+$ and
$h_{\times}$ are independent of $r$ at large $r$. 
We determine $h_+$ and $h_{\times}$ from the Moncrief~\cite{m74} variables
$G_{lm}$, $h_{1lm}$, $H_{2lm}$ and $K_{lm}$, which we compute by
quadratures of our metric over a spherical shell near the edge of our
grid (see~\cite{wextract}).  From these we find the gauge-invariant
linear combinations $R^{\rm E}_{lm}$.  In our code, we work with real,
orthonormal combinations of spherical harmonics, and so we compute the
amplitudes $R_{22+}$, which is associated with the mode having angular
dependence $(Y_{22}+Y_{2-2})/\sqrt{2}$, and $R_{22-}$, which is
associated with the mode having angular dependence
$(Y_{22}-Y_{2-2})/i\sqrt{2}$.


As argued in \cite{wextract}, our gauge conditions approach the TT
gauge as $r$ becomes large.  In this gauge, the two gravitational wave
polarizations can be constructed from
\begin{equation}
\label{polars1}
h_+ = r \sum_{lm} G_{lm} W_{lm} \qquad
h_{\times} = r \sum_{lm} G_{lm} X_{lm}/\sin\theta.
\end{equation}
The Moncrief functions only depend on $r$ and $t$, while the angular
dependence is contained in the $W_{lm}$ and $X_{lm}$.  In the TT
gauge, all Moncrief variables vanish except for the $G_{lm}$, so that
the linear combination for the $R^{\rm E}_{lm}$ reduces to
\begin{equation}
\label{Mvars}
R^{\rm E}_{\rm 2\pm 2} = 2\sqrt{12} \, G_{\rm 2\pm 2}
\end{equation}
for the dominant modes $l = 2$, $m = \pm 2$.  Inserting these into
eq.~(\ref{polars1}) we then construct
\begin{equation}
\label{polars2}
\begin{array}{rcl}
 h_+ &=& \displaystyle
	{r\over 2}\sqrt{5\over 16\pi}
	(1+\cos^2\theta)( \cos(2\phi) R_{\rm 22+} + 
	\sin(2\phi) R_{\rm 22-} ) \\[3mm]
   h_{\times} 
	&=& \displaystyle 
	r\sqrt{{5\over 16\pi}}\cos\theta 
	(-\sin(2\phi)R_{\rm 22+} + \cos(2\phi)R_{\rm 22-}) 
\end{array}
\end{equation}
up to an arbitrary phase.  Finally, the amplitude $A$ on the $z$-axis
(i.e.~$\theta = 0$) is given by
\begin{equation}
\label{AvsR}
A = r \sqrt{{5\over 16\pi}} \sqrt{(R_{\rm 22+})^2 + (R_{\rm 22-})^2}.
\end{equation}
As a check of the above, one can, by making explicit the time
dependence in (\ref{polars2}) ($\phi\to\phi - \Omega_{\rm orb}t$), use
Eq.~(\ref{dEvsR}) to derive the luminosity-amplitude relationship for
quadrupole (e.g. point mass) systems, Eq.~(\ref{Apn}).  The amplitude
of the waveform $h(t)$ in Eq.~(\ref{h_wt}) is then equal to $A(t)/r$. 
(The $1/r$ falloff is restored.)

\end{appendix}


\begin{references}

\bibitem{tetal01} H. Tagoshi {\em et.al.}, Phys. Rev. D {\bf 63},
	062001 (2001). 

\bibitem{t98} K. S. Thorne, in
	{\it Black Holes and Relativistic Stars},
	edited by R.M. Wald (U. of Chicago Press, Chicago, 1998),
	p. 59.  

\bibitem{bdiww95} L. Blanchet, T. Damour, B.R. Iyer, C.M. Will, and 
	A.G. Wiseman, Phys. Rev. Lett {\bf 74} 3515 (1995). 

\bibitem{dis98} T. Damour, B. R. Iyer, and B. S. Sathyaprakash,
	Phys. Rev. D {\bf 57}, 885 (1998).

\bibitem{bd99} A. Buonanno, and T. Damour, Phys. Rev. D {\bf 59}, 
	084006 (1999).

\bibitem{djs00} T. Damour, P. Jaranowski and G. Sch\"afer, 
	Phys. Rev. D {\bf 62}, 084011 (2000).

\bibitem{su00} M. Shibata and K. Ury\=u, Phys. Rev. D {\bf 61}, 
	064001 (2000).

\bibitem{pp94} R. H. Price, J. Pullin, Phys. Rev. Lett. {\bf 72}, 3297 (1994).

\bibitem{ketal99} G. Khanna {\em et. al.}, Phys. Rev. Lett. {\bf 83}, 3581
	(1999).

\bibitem{bct98} P. R. Brady, J. D. E. Creighton, and K. S. 
	Thorne, Phys. Rev. D {\bf 58}, 061501 (1998).

\bibitem{dbs01} M. D. Duez, T. W. Baumgarte, and S. L. Shapiro,
	Phys. Rev. D {\bf 63}, 084030 (2001).

\bibitem{bcsst97} T. W. Baumgarte, G. B. Cook, M. A. Scheel, S. L. Shapiro
	and S. A. Teukolsky, Phys. Rev. Lett. {\bf 79}, 1182 (1997);
	Phys. Rev. D {\bf 57}, 7299 (1998).

\bibitem{ue00} K. Ury\=u and Y. Eriguchi, Phys. Rev. D {\bf 61}, 
	124023 (2000).

\bibitem{bhs99} T. W. Baumgarte, S. A. Hughes, and S. L. Shapiro, 
	Phys. Rev. D {\bf 60}, 087501 (1999).

\bibitem{ybs01} H.-J. Yo, T. W. Baumgarte and S. L. Shapiro, 
	Phys. Rev. D. {\bf 63}, 064035 (2001).

\bibitem{irr_1} C. S. Kochanek, Astrophys. J. {\bf 398} 234 (1992);
	L. Bildsten and C. Cutler, Astrophys. J. {\bf 400}, 175 (1992).

\bibitem{bs99} T. W. Baumgarte and S. L. Shapiro, Phys. Rev. D {\bf 59}, 
	024007 (1999).

\bibitem{sn95} M. Shibata and T. Nakamura, Phys. Rev. D {\bf 52}, 5428
	(1995).

\bibitem{use00} K. Ury\=u, M. Shibata, and Y. Eriguchi, Phys. Rev. D
	{\bf 62} 104015 (2000).

\bibitem{w94} C. M. Will, in {\it Relativistic Cosmology}, 
	edited by M. Sasaki (Universal Academy Press,1994), p. 95.

\bibitem{ab99} W. D. Arnett and R. L. Bowers, Astrophys. J. Suppl. 
	{\bf 33}, 415 (1977).

\bibitem{m74} V. Moncrief, Ann. Phys. {\bf 88}, 323 (1974).

\bibitem{lrs94} D. Lai, F. A. Rasio, and S. L. Shapiro,
	Astrophys. J. {\bf 420}, 811 (1994).

\bibitem{b01} T. W. Baumgarte, to appear in Astrophysical Sources of
	Gravitational Radiation, edited by J. M. Centrella, AIP
	Conference Proceedings, New York (2001) (also gr-qc/0101045).

\bibitem{note} Setting the maximum rest-mass of an isolated neutron star
	equal to about 1.9 $M_{\odot}$ ~\cite{apr98} implies that the stars
	in our simulations have rest masses of about 1.5 $M_{\odot}$.

\bibitem{apr98} A. Akmal, V. R. Pandharipande, and D. G. Ravenhall,
	Phys. Rev. C. {\bf 58}, 1804-1828 (1998).

\bibitem{note2} In generating the merger waveform in~\cite{su00},
	only approximate QE initial data was used.  This induces
	some inaccuracy in the merger phase of the waveform shown
	in Figure 6.  Accordingly, Figure 6 should not be regarded
	as definitive but, rather, illustrative of the technique and
	the general shape of the expected waveform.

\bibitem{cabffkmopst93} C. Cutler, T. A. Apostolatos, L. Bildsten, L. S. Finn,
	E. E. Flanagan, D. Kennefick, D. M. Markovic, A. Ori, E. Poisson,
	G. J. Sussman, and K. S. Thorne, Phys. Rev. Lett. {\bf 70},
	2984 (1993).

\bibitem{outer_BC} N. T. Bishop, R. Gomez, P. R. Holvorcem,
	R. A. Matzner, P.  Papadopoulos and J. Winicour,
	J. Comp. Phys. {\bf 136}, 140 (1997); A. M. Abrahams {\it
	et.al.} (The Binary Black Hole Grand Challenge Alliance)
	Phys. Rev. Lett. {\bf 80}, 1812 (1998).

\bibitem{wextract} M. Shibata, Prog. Theor. Phys. {\bf 101}, 1199 (1999).

\end{references}
\end{document}